\documentclass[12pt, a4paper]{article}
\usepackage{amssymb}
\usepackage{amsmath}
\usepackage{graphicx}
\usepackage{subfigure}

\usepackage[authoryear]{natbib}
\usepackage{times,epsfig,makeidx,amsfonts,amsmath,dcolumn,enumerate,multirow,graphicx,amssymb,bm,color}
\usepackage[margin=1in]{geometry}
\usepackage{algorithm}
\usepackage[]{algorithmic}
\usepackage{setspace}

\newtheorem{theorem}{Theorem}

\newenvironment{proof}[1][Proof]{\noindent\textbf{#1.} }{\ \rule{0.5em}{0.5em}}

\begin{document}
\title{Nonparametric inference for $P(X<Y)$ with paired variables}
\author{  {\large Jos{\'e} Arturo Montoya}\footnote{{\sc Universidad de Sonora, Departamento de Matem{\'a}ticas.} E-mail:  montoya@mat.uson.mx}  \,{\large and} {\large Francisco Javier Rubio}\footnote{{\sc University of Warwick, Department of Statistics, Coventry, CV4 7AL, UK.} E-mail:  F.J.Rubio@warwick.ac.uk}\\
}
\date{}
\maketitle

\begin{abstract}
We propose two classes of nonparametric point estimators of $\theta=P(X<Y)$ in the case where $(X,Y)$ are paired, possibly dependent, absolutely continuous random variables. The proposed estimators are based on nonparametric estimators of the joint density of $(X,Y)$ and the distribution function of $Z=Y - X$. We explore the use of several density and distribution function estimators and characterise the convergence of the resulting estimators of $\theta$. We consider the use of bootstrap methods to obtain confidence intervals. The performance of these estimators is illustrated using simulated and real data. These examples show that not accounting for pairing and dependence may lead to erroneous conclusions about the relationship between $X$ and $Y$.
\end{abstract}

\noindent {\it Key Words: Bootstrap, dependence, density estimation, distribution estimation, stress-strength model.}

\section{Introduction}

The study of stress--strength models have received considerable attention for many years due to its applicability in diverse areas. The main interest in this kind of models is the quantity $\theta=P(X<Y)$, where $X$ and $Y$ are random variables. In medicine for example, if $X$ and $Y$ are the outcomes of a control and an experimental treatment respectively, the parameter $\theta$ can be interpreted as the effectiveness of treatment $Y$ \citep{V11}. This quantity is also related to the Receiver Operating Characteristic (ROC) curves, where $\theta$ is interpreted as an index of accuracy \citep{Z08}. In engineering and reliability studies $\theta$ is also a quantity of interest because it may represent the probability that the strength of a component ($Y$) exceeds the stress ($X$) coming from external factors \citep{K03}.

Stress-strength models were introduced by \cite{B56} who proposed a nonparametric estimator of $\theta$ based on the Mann-Whitney statistic for the case where $X$ and $Y$ are independent. There is a large amount of literature related to the study of point and interval estimation of $\theta$ using different approaches (see \citealp{K03} for a good survey on this). For instance, in the case where $X$ and $Y$ are independent, \cite{S98} proposes a Bayesian approach using reference priors; \cite{B06} propose an estimator based on kernel estimators of the densities of $X$ and $Y$ (which can be straightforwardly generalised to the use of other nonparametric density estimators); \cite{Z08} proposes the use of bootstrap and asymptotic intervals; \cite{JYZ09} estimate $\theta$ using the empirical likelihood; \cite{M08} and \cite{DM12} propose the use of the profile likelihood for conducting inference about $\theta$; and \cite{V11} propose the use of Bayesian inference with Jeffreys and matching priors as well as modified profile likelihoods for the cases where $X$ and $Y$ are normal or exponential random variables.

It is important to mention that the parameter $\theta$ may not be available in a closed form in many cases (see \citealp{AC04} and \citealp{GB01} for an example of this). This makes difficult (if at all feasible) to find a reparameterisation involving $\theta$, which complicates the use of the classical approach. In particular, the use of the profile likelihood might be difficult if this reparameterisation is not available \citep{DM12}. Alternative inferential approaches that overcome this difficulty are Bayesian inference, nonparametric estimation, and the use of bootstrap methods, which allow for obtaining confidence and credible intervals for the parameter of interest \citep{B06,Z08,RS12}.

New interest has been focused on the estimation of $\theta$ in the case where $X$ and $Y$ are dependent random variables. For example \cite{B11} assumes that $(X,Y)$ are jointly normally distributed; \cite{RS12} suppose that $X$ and $Y$ are marginally distributed as skewed scale mixture of normals and construct the corresponding joint distribution using a Gaussian Copula; \cite{DG12a} construct the joint distribution of $(X,Y)$ using a Farlie-Gumbel-Morgenstern copula with marginal distributions belonging to the Burr system; \cite{DG12b} consider Dagum distributed marginals and construct their joint distribution using a Frank copula; among others \citep{N05,GGA12}. In these papers, the importance of taking the assumption of dependence between $X$ and $Y$ into consideration is illustrated using simulated and real data sets.

We propose two classes of nonparametric estimators of $\theta$ for the case where $(X,Y)$ are paired, possibly dependent, continuous random variables. This scenario is of interest since paired observations are produced in many experimental designs (see e.g. \citealp{S00} and \citealp{CR00} for examples of this). The estimators proposed here are based on nonparametric estimators of the density of $(X,Y)$ and the distribution function of $Z=Y-X$. This approach avoids making distributional assumptions over $(X,Y)$ and allows for interval estimation of $\theta$ via nonparametric bootstrap. In addition, this method can be easily implemented in R using already existing packages. In Section \ref{NonparametricEstimators} we introduce these estimators and prove some asymptotic properties for the choice of several nonparametric estimators. We also detail how to combine kernel density estimation (KDE) with the methods proposed here. In Section \ref{Example} we present two examples, using simulated and real data, which illustrate the importance of accounting for pairing and dependence of the observations when conducting inference about $\theta$.


\section{Nonparametric estimators of $\theta$}\label{NonparametricEstimators}

Let $(X,Y)$ be a pair of absolutely continuous random variables with joint density $f_{X,Y}:{\mathbb R}^2\rightarrow {\mathbb R}_+$. By definition, we have that

\begin{eqnarray}\label{estdens}
\theta = {\mathbb P}(X<Y) = \int_{-\infty}^{\infty}\int_{-\infty}^y f_{X,Y}(x,y)dxdy.
\end{eqnarray}

Alternatively, by defining the variable $Z=Y-X$ we obtain

\begin{eqnarray}\label{estdist}
\theta = {\mathbb P}(Z>0)=\int_0^{\infty}f_Z(z)dz=1-F_Z(0)=S_Z(0),
\end{eqnarray}

\noindent where $f_Z$, $F_Z$ and $S_Z$ are the density function, the cumulative distribution function, and the survival function of $Z$, respectively. These equivalent expressions suggest the following nonparametric methods for estimating the parameter $\theta$.

\subsection{Estimator I}\label{Estimator I}

Let $({\bf x}, {\bf y})$ be a sample from $(X,Y)$ of size $n$ and suppose that these observations are collected in couples $(x_i,y_i)$, $i=1,\dots,n$. The first proposed estimator, based on expression (\ref{estdens}), consists of substituting the density $f_{X,Y}$ by a nonparametric density estimator as follows.

\begin{algorithm}
\caption{}\label{alg:depden}
\begin{algorithmic}[1]
\STATE Using the sample $({\bf x}, {\bf y})$ construct a nonparametric estimator $\hat{f}_{X,Y}$ of the density $f_{X,Y}$.
\STATE Define the estimator $\tilde\theta = \int_{-\infty}^{\infty}\int_{-\infty}^y \hat{f}_{X,Y}(x,y)dxdy$.
\end{algorithmic}
\end{algorithm}

Note that Algorithm \ref{alg:depden} involves both a two-dimensional density estimation and the calculation of a double integral. Several nonparametric density estimators can be employed for this purpose such as kernel density estimators \citep{P62}, shape-restricted estimators \citep{C10}, among others \citep{S92}. This choice has, of course, implications on the performance of the estimators. In Section \ref{AsymptoticProperties} we present some asymptotic properties of $\tilde\theta$ for different choices of $\hat{f}_{X,Y}$. The integration step can be conducted using quadrature or Monte Carlo methods.

\subsection{Estimator II}\label{Estimator II}

Again, let $({\bf x}, {\bf y})$ be a sample from $(X,Y)$ of size $n$ and suppose that these observations are collected in couples $(x_i,y_i)$, $i=1,\dots,n$. Define the vector of differences ${\bf z}={\bf y}-{\bf x}$. The second proposed estimator, based on expression (\ref{estdist}), is constructed as follows.

\begin{algorithm}
\caption{}\label{alg:depdist}
\begin{algorithmic}[1]
\STATE Calculate the differences ${\bf z}={\bf y}-{\bf x}$.
\STATE Using the sample ${\bf z}$ construct a nonparametric estimator $\hat{F}_Z$ of the distribution function of $Z$ .
\STATE Define the estimator $\hat\theta = 1-\hat F_Z(0)$.
\end{algorithmic}
\end{algorithm}

For the nonparametric distribution estimator $\hat{F}_Z$ in step $2$ we can employ the empirical cumulative distribution function (ECDF) or the induced distribution estimators obtained by integrating a nonparametric density estimator $\hat{f}_Z$, which lead to $\hat{\theta} = \int_0^{\infty}\hat{f}_Z(z)dz$. In this line, several univariate nonparametric estimators of $\hat{f}_Z$ can be considered such as kernel density estimators \citep{P62}, shape-restricted density estimators \citep{C10} and smooth shape-restricted estimators \citep{DR09,C10,DR11}.

Note that the use of both, Estimator I and Estimator II, avoids making assumptions on the distribution of $(X,Y)$ and the sort of dependence between the variables $X$ and $Y$. The relationship between these variables, which can be either dependent or independent, is implicitly included in the nonparametric estimators of the density (distribution). In addition, the use of nonparametric bootstrap coupled with either Algorithm \ref{alg:depden} or Algorithm \ref{alg:depdist} allows for obtaining a variety of bootstrap confidence intervals for these estimators \citep{DE96}.

\subsection{Use of Estimator I and Estimator II with kernel density estimators}\label{E1E2KDE}

In this section we discuss the use of KDE in Algorithms I and II. Recall that the use of KDE involve the choice of two elements: a kernel function and a bandwidth parameter (or bandwidth matrix in a multivariate framework). Here, we present a brief discussion on appropriate choices for these elements in our context.

\subsubsection{Estimator I}

Let $H$ be a symmetric, positive definite, $2\times 2$ bandwidth matrix and $k_2$ be a two-dimensional kernel function \citep{P62}. Define also $k_H({\bf t})=(\det H)^{-\frac{1}{2}} k_2(H^{-\frac{1}{2}}{\bf t})$, ${\bf t}\in{\mathbb R}^2$. If we consider the use of a KDE in step 1 of Algorithm \ref{alg:depden}, then the estimator $\tilde\theta$ can be written as

\begin{eqnarray}\label{kdfest}
\tilde\theta = \dfrac{1}{n}\sum_{j=1}^n \int_{-\infty}^\infty\int_{-\infty}^y k_H(x-x_j,y-y_j)dxdy,
\end{eqnarray}

\noindent which can be calculated using quadrature or Monte Carlo methods. The implementation of this estimator requires the specification of the kernel function $k_2$ and the bandwidth matrix $H$. A natural first choice is the use of a bivariate Gaussian kernel $\phi_2=k_2$. The choice of the bandwidth matrix $H$ can be crucial for the performance of KDE, which has fostered an extensive study of several bandwidth matrix estimators (see \citealp{DH05} for a good survey on this). However, appropriate bandwidth matrices for estimating the distribution involved in (\ref{kdfest}) seem to have been little studied to our knowledge. Nevertheless, as a first approach one can consider bandwidth matrix estimators employed in KDE such as the plug-in and cross-validation bandwidth estimators, which are implemented in the R package `ks' \citep{D11}.

\subsubsection{Estimator II}

Let $k_1$ be a one--dimensional kernel and $h>0$ be the corresponding bandwidth (also termed \emph{smoothing parameter}). If we consider the use of a univariate KDE in step 2 of Algorithm \ref{alg:depdist}, then the estimator $\hat\theta$ can be written as

\begin{eqnarray}\label{distkerest}
\hat\theta = \dfrac{1}{nh}\sum_{j=1}^n \int_{0}^{\infty} k_1\left(\dfrac{z-z_j}{h}\right)dz = 1- \dfrac{1}{n}\sum_{j=1}^n K_1\left(\dfrac{z_j}{h}\right),
\end{eqnarray}

\noindent where $K_1\left(\dfrac{z}{h}\right)=\dfrac{1}{h}\int_{0}^{\infty} k_1\left(\dfrac{z}{h}\right)dz$. Again, a natural first choice is the Gaussian kernel $\Phi=K_1$. The choice of the bandwidth $h$ in the context of density estimation has been extensively studied, we refer the reader to \cite{J96} for a good survey on this. However, the choice of this parameter in the context of kernel distribution function estimation has received less attention. \cite{QP12} present a compendium of appropriate bandwidth parameters in the context of kernel distribution estimator, they also implement these in the R package `kerdiest'.



\subsection{Results on the convergence of the proposed estimators}\label{AsymptoticProperties}

The convergence of Estimator I coupled with KDE is difficult to assess given the limited literature about the choice of appropriate bandwidth matrices for estimating a bivariate distribution. Despite this limitation, one can expect a good performance of this estimator for moderate or large samples and the use of any reasonable bandwidth matrix since kernel estimators converge in terms of the mean square and mean absolute errors to the true density. The following result shows that, even using a diagonal bandwidth matrix, the resulting estimator of $\theta$ is weakly consistent under rather mild conditions. The use of more appropriate bandwidth matrices is therefore expected to produce better estimators.

\begin{theorem}\label{T1}
Suppose that $k_2$ is bounded on ${\mathbb R}^2$ with

\begin{eqnarray*}
L(u)=\sup_{\vert\vert {\bf t} \vert\vert\geq u }k_2({\bf{t}}),
\end{eqnarray*}

\noindent for $u\geq 0$. Let $\{h_n\}_{n=1}^{\infty}$ be a sequence of positive numbers such that $\lim_{n\rightarrow\infty}h_n=0$ and $\lim_{n\rightarrow\infty}n h_n^2=\infty$. Define the sequence of bandwidth matrices $H_n=\operatorname{diag}(h_n)$. Suppose also that one of the following conditions holds

\begin{enumerate}[(i)]
\item $\vert\vert {\bf t} \vert\vert^2 k_2({\bf t}) \rightarrow 0$ as $\vert\vert {\bf t} \vert\vert \rightarrow \infty$ and $f_{X,Y}$ is almost surely continuous.
\item $f_{X,Y}$ is bounded.
\item $\int_0^{\infty} uL(u)du <\infty$.
\end{enumerate}

Then, $\tilde{\theta}$ is a weakly consistent estimator of $\theta$, this is, $\tilde{\theta}\stackrel{{\mathbb P}}{\rightarrow}\theta$, as $n\rightarrow\infty$.

\proof First, note that
\begin{eqnarray*}
\vert \tilde{\theta}-\theta\vert &=& \left\vert \int_{-\infty}^{\infty}\int_{-\infty}^y [\hat{f}_{X,Y}(s,t) - f_{X,Y}(s,t)] ds dt \right\vert  \\
&\leq&  \int_{-\infty}^{\infty}\int_{-\infty}^y \left\vert\hat{f}_{X,Y}(s,t) - f_{X,Y}(s,t)\right\vert ds dt  \\
&\leq&  \int_{-\infty}^{\infty}\int_{-\infty}^{\infty} \left\vert\hat{f}_{X,Y}(s,t) - f_{X,Y}(s,t)\right\vert ds dt  \\
&\leq& \operatorname{MAE}(\hat{f}_{X,Y} ,f_{X,Y}),
\end{eqnarray*}

\noindent where $\operatorname{MAE}$ denotes the \emph{mean absolute error} which is also the $L_1$ distance. Under the stated assumptions we have that $\lim_{n\rightarrow\infty}\operatorname{MAE}(\hat{f}_{X,Y} ,f_{X,Y})= 0$, in probability, by the Theorem in \cite{DW79}.
\end{theorem}

Although the use of the shape-restricted density estimator in \cite{C10} does not involve a tuning parameter, a study of the asymptotic properties of the induced distribution estimator seems not to have been done yet. However, since this density estimator has smaller mean integrated squared error than those obtained with KDE methods \citep{C10}, the use of this method in Algorithm \ref{alg:depden} is also expected to produce good estimators of $\theta$ for moderate or large samples.

On the other hand, given the immediate relationship between the Estimator II and the estimation of the distribution $F_Z$, it follows that the asymptotic properties of $\hat\theta$ are inherited from those of the estimator $\hat{F}_Z$ evaluated at $0$. Some specific asymptotic results are presented below for different estimators of $\hat{F}_Z(0)$.

The following result shows that the use of the empirical distribution for estimating $\hat{F}_Z(0)$ produces consistent and asymptotically normal estimators of $\theta$.

\begin{theorem}\label{T2}
Let ${\hat F}_Z$ be the empirical distribution function, then
\begin{enumerate}[(i)]
\item $\hat{\theta}$ is strongly consistent, this is, $\hat{\theta}\stackrel{a.s.}{\rightarrow}\theta$, as $n\rightarrow\infty$.
\item The estimator $\hat{\theta}$ is asymptotically normal, this is
\begin{eqnarray*}
\sqrt{n}\left(\hat{\theta}-\theta\right)\stackrel{d}{\rightarrow} N\left(0,F_Z(0)(1-F_Z(0))\right),
\end{eqnarray*}
\noindent as $n\rightarrow\infty$.
\end{enumerate}
\proof The results follow by the law of large numbers and the central limit theorem \citep{v98}.
\end{theorem}

The use of kernel distribution estimators can also produce consistent and asymptotically normal estimators of $\theta$ under certain conditions as indicated in the following theorem.

\begin{theorem}\label{T3}
Assume that $F_Z$ is uniformly Lipschitz on ${\mathbb R}$ and let ${\hat F}_Z$ be a \emph{regular} kernel estimator. This is, there exists a positive sequence $\{h_n\}_{n=1}^{\infty}$ such that $h_n=o\left(n^{-\frac{1}{2}}\right)$ and
\begin{eqnarray*}
\int_{\vert t \vert >h_n}  \dfrac{1}{h_n} k_1\left(\dfrac{t}{h_n}\right) dt = o\left(n^{-\frac{1}{2}}\right).
\end{eqnarray*}
Then, it follows that

\begin{enumerate}
\item $\hat{\theta}$ is a strongly consistent estimator of $\theta$.
\item $\hat{\theta}$ is asymptotically normal, $\sqrt{n}\left(\hat{\theta}-\theta\right)\stackrel{d}{\rightarrow} N\left(0,F_Z(0)(1-F_Z(0))\right)$.
\end{enumerate}
\proof
(i) Using the triangle inequality it follows that $\vert \hat{F}_Z(0) - F_Z(0)\vert \leq \vert \hat{F}_Z(0) - F_n(0)\vert + \vert \hat{F}_n(0) - F_Z(0)\vert$, where $F_n$ is the empirical distribution function. Then, the result follows by Theorem 2.3 from \cite{F91} and the law of large numbers. (ii) The asymptotic normallity of $\hat{\theta}$ follows by Corollary 2.4 from \cite{F91}.
\end{theorem}

By relaxing the assumptions of the previous theorem it is possible to prove that the use of kernel distribution estimators also produces weakly consistent estimators of $\theta$.

\begin{theorem}\label{T4}
Suppose that $k_1$ is bounded in ${\mathbb R}$ with

\begin{eqnarray*}
L(u)=\sup_{\vert t \vert\geq u }k_1(t),
\end{eqnarray*}

\noindent for $u\geq 0$. Let $\{h_n\}_{n=1}^{\infty}$ be a sequence of positive bandwidths such that $\lim_{n\rightarrow\infty}h_n=0$ and $\lim_{n\rightarrow\infty}n h_n=\infty$. Suppose also that one of the following conditions holds

\begin{enumerate}[(i)]
\item $\vert t \vert k_1(t) \rightarrow 0$ as $\vert t \vert \rightarrow \infty$ and $f_Z$ is almost surely continuous.
\item $f_Z$ is bounded.
\item $\int_0^{\infty} L(u)du <\infty$.
\end{enumerate}

Then, $\hat{\theta}$ is a weakly consistent estimator of $\theta$.

\proof First, note that
\begin{eqnarray*}
\vert\hat{\theta}-\theta\vert &=& \left\vert \hat{F}_Z(0)-F_Z(0)\right\vert =  \left\vert \int_{-\infty}^0 [\hat{f}_Z(t)-f_Z(t)] dt \right\vert \\
&\leq&  \int_{-\infty}^0 \left\vert \hat{f}_Z(t)-f_Z(t)\right\vert dt \\
&\leq&  \int_{-\infty}^{\infty} \left\vert \hat{f}_Z(t)-f_Z(t)\right\vert dt  \\
&\leq& \operatorname{MAE}(\hat{f}_Z ,f_Z),
\end{eqnarray*}

\noindent where $\operatorname{MAE}$ denotes the \emph{mean absolute error}. Under the stated assumptions we have that $\lim_{n\rightarrow\infty}\operatorname{MAE}(\hat{f}_Z ,f_Z)= 0$, in probability, by the Theorem in \cite{DW79}.
\end{theorem}

Note that Theorems \ref{T1} and \ref{T4} simply require a well-behaved kernel function and the boundedness of the target density. The assumptions on the bandwidth parameters are also rather mild since most of the popular choices satisfy these conditions.

The use of shape-restricted estimators, by their nature itself, require additional assumptions on the target density. The following result presents such conditions that produce consistent estimators of $\theta$.

\begin{theorem}\label{T5}
Let $\hat{F}_Z$ be the shape-restricted nonparametric estimator of $F_Z$ proposed in \cite{DR09} and suppose that the log-density $\log (f_Z)$ is Lipschitz continuous and $\log (f_Z)^{\prime}$ is H{\"o}lder continuous of order $\beta\in[1,2]$ on a compact interval $I\subset {\mathbb R}$. Then, $\hat{\theta}$ is a weakly consistent estimator of $\theta$.

\proof The result is a consequence of Corollary 4.2 from \cite{DR09}.
\end{theorem}

The results presented in this section show that both estimators have good asymptotic performance under mild conditions. An important difference between Estimator I and Estimator II is that the former involves a two-dimensional density (distribution) estimation while the latter involves a one-dimensional density (distribution) estimation. This represents an advantage of Estimator II over Estimator I since the convergence rate of the resulting estimators as well as the ease of implementation is tied to the dimensionality of the problem. However, an interesting feature of Estimator I is that it can be implemented in the context of censored and missing observations since the use of KDE in these contexts has been studied, for example, in \cite{TM83} and \cite{WY96}.


\section{Examples}\label{Example}

In this section, we present two examples that illustrate the implementation of the estimators proposed in Section \ref{NonparametricEstimators}. In the first example we use a sample simulated from a bivariate sinh-arcsinh distribution \citep{JP09}. As detailed in \cite{JP09}, this distribution contains parameters that control skewness, kurtosis and correlation of the marginals. This example illustrates the influence of the assumptions of pairing and dependence on the bootstrap distributions of the corresponding estimators in terms of their location and spread. In the second example we use a real data set and show that not including the assumptions of pairing and dependence may lead to opposite conclusions about the relationship between $X$ and $Y$. In both examples, we consider the following 9 types of estimators of $\theta$:

\begin{enumerate}[(i)]

\item {\bf Kernel 2D.} Based on Algorithm \ref{alg:depden}, this estimator employs a two-dimensional Gaussian kernel density estimator with the bandwidth matrix \emph{Hscv} implemented in the R package `ks' \citep{D11}. The required integration step is conducted using quadrature methods.

\item {\bf MLE 2D.} This estimator is based on Algorithm \ref{alg:depden}. The corresponding two-dimensional density estimation is conducted using the shape-restricted estimator from \cite{C10}. This estimator is also implemented using the command \emph{dlcd} from the R package `LogconcDEAD' \citep{CGS09}. The integration of this density is conducted using a Monte Carlo method.

\item {\bf SMLE 2D.} This estimator is based on Algorithm \ref{alg:depden}. The corresponding two-dimensional density estimation is conducted using the smooth shape-restricted estimator \citep{C10} implemented in the command \emph{dslcd} from the R package `LogconcDEAD' \citep{CGS09}. The integration of this density is conducted using a Monte Carlo method.
\item {\bf MLE 1D.} Estimates $\hat{F}_Z$ in Algorithm \ref{alg:depdist} by integrating the shape-restricted density estimator proposed in \cite{C10}. The density estimation is implemented using the command \emph{dlcd} from the R package `LogconcDEAD' \citep{CGS09}.

\item {\bf Kernel 1D.} Based on Algorithm \ref{alg:depdist}, this estimator employs a Gaussian kernel distribution estimator with the bandwidth \emph{ALbw} implemented in the R package `kerdiest' \citep{QP12}.

\item {\bf SMLE 1D.} Estimates $\hat{F}_Z$ in Algorithm \ref{alg:depdist} by integrating a smoothed shape-restricted density estimator \citep{C10} implemented in the command \emph{dslcd} from the R package `LogconcDEAD' \citep{CGS09}.

\item {\bf ECDF.}  This estimator employs the empirical distribution function for estimating $\hat{F}_Z$ in Algorithm \ref{alg:depdist}.

In order to assess the impact of the assumptions of pairing and dependence in the estimation of $\theta$, we also consider the following estimators:

\item {\bf Independent.} \citep{B06} This estimator assumes that $X$ and $Y$ are independent variables and that the corresponding samples are unpaired. The estimator is defined as

\begin{eqnarray}\label{indestimator}
\theta^{\star} = \int_{-\infty}^{\infty}\int_{-\infty}^y \hat{f}_X(x)\hat{f}_Y(y)dxdy,
\end{eqnarray}

\noindent where  $\hat{f}_X$ and $\hat{f}_Y$ are Gaussian kernel density estimators obtained with samples of $X$ and $Y$ respectively. For both KDE we employ the bandwidth $h=\left(\dfrac{4\hat{\sigma}^5}{3n}\right)^{\frac{1}{5}}$, where $\hat{\sigma}$ is the sample standard deviation and $n$ is the sample size. This bandwidth is known as the Silverman's rule of thumb.

\item {\bf Paired.} \citep{B06}. This estimator is the same as (\ref{indestimator}) but assuming that the samples of $X$ and $Y$ are paired. This additional assumption is taken into consideration in the bootstrap methods used to calculate confidence intervals for $\theta^{\star}$.

\end{enumerate}

Bootstrap samples and bootstrap confidence intervals (Normal, Basic, Percentile and BCa) are obtained using the R packages `boot' \citep{CR12} and `simpleboot' \citep{P08}. R source code for these examples is available upon request.

\subsection{Simulated data}

\begin{figure}[htp]
\begin{center}
\begin{tabular}{c c}
\psfig{figure=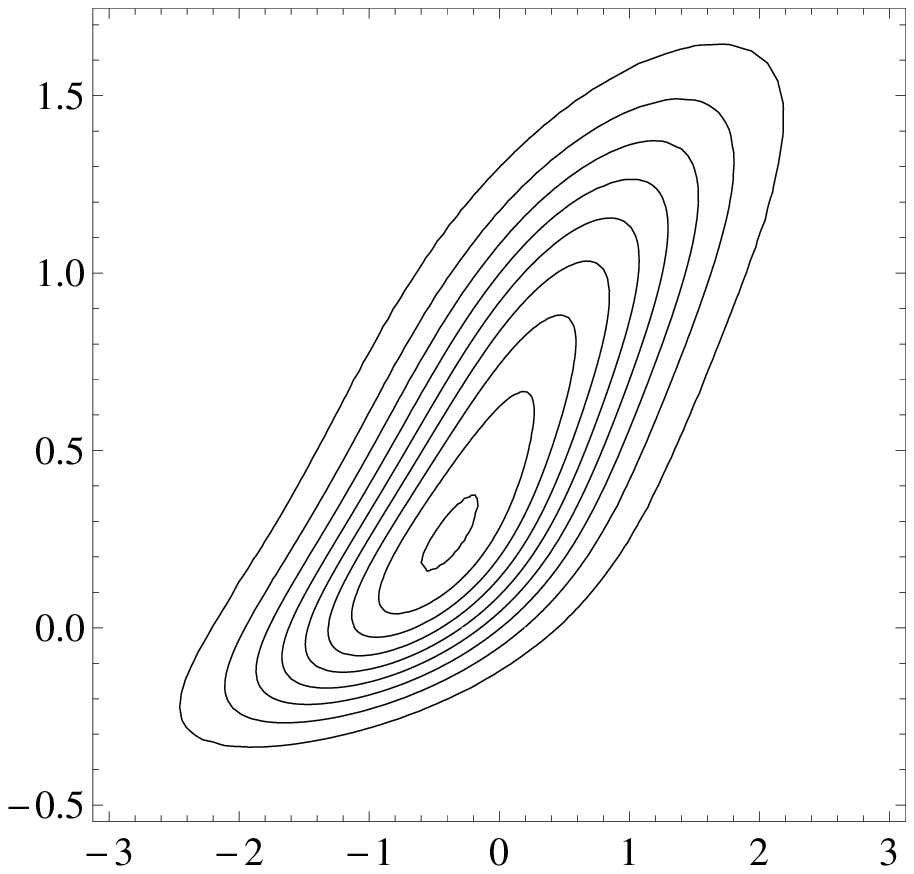,  height=5cm}  &
\psfig{figure=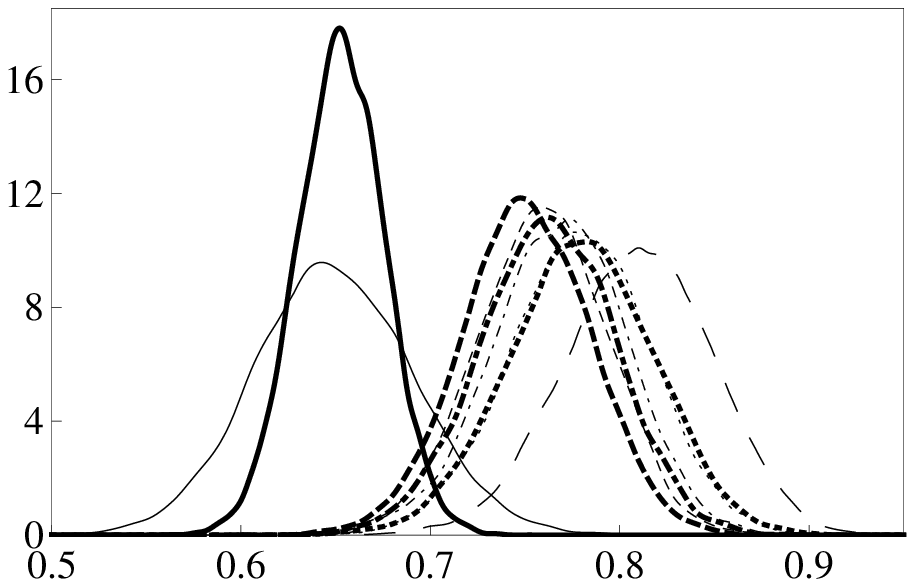,  height=5cm}\\
(a) & (b)
\end{tabular}
\end{center}
\caption{\small (a) Contour plot: sinh-arcsinh distribution; (b) Simulated data: bootstrap distributions of the estimators. Independent (solid line), Paired (solid bold line), ECDF (long-dashed line), Kernel 1D (dashed line), Kernel 2D (dashed bold line), MLE 1D (dotted line), MLE 2D (dotted bold line), SMLE 1D (dotted-dashed line), SMLE 2D (dotted-dashed bold line).}
\label{fig:contoursas}
\end{figure}

In this example we use a simulated sample of size $n=100$ from a bivariate sinh-arcsinh distribution \citep{JP09} with parameters $(\sigma_1,\sigma_2,\rho,\epsilon_1,\epsilon_2,\delta_1,\delta_2)=(1,1,0.75,0,1,1,2)$. Figure \ref{fig:contoursas}a shows a contour plot of the corresponding density. This is a complex scenario where the entries present departure from normality and dependence. The population correlation coefficient of this sample is $0.737$ and the theoretical correlation is $0.743$. The parameter $\theta$ in this family of distributions is not generally tractable. The theoretical value of $\theta$, obtained by numerical integration, is $0.78$. Figure \ref{fig:contoursas}b shows the bootstrap distribution of the estimators of $\theta$ previously described. We can observe a considerable influence of the assumptions of pairing and dependence in the location and spread of the bootstrap distributions of the estimators of $\theta$. We can also notice the influence of these assumptions in the point estimators and bootstrap confidence intervals shown in Table \ref{table:EstCISAS}. In this case, not including these assumptions leads to underestimating $\theta$. Finally, we can observe that the estimator ECDF is slightly larger than the others, which seems to be a result of its discrete nature.

\begin{table}[h!]
\begin{center}
\begin{tabular}[h]{ c c c c c c }
\hline
Estimator & $\hat{\theta}$ & Normal & Basic & Percentile & BCa \\
\hline
Independent & 0.65  & $( 0.560,  0.724  )$ & $(  0.559,  0.723 )$ & $( 0.568,  0.732   )$ & $( 0.562,  0.727  )$ \\
\hline
Paired & 0.65  & $( 0.606,  0.695  )$ & $(  0.606,  0.696 )$ & $(  0.607,  0.697 )$ & $( 0.604,  0.694  )$ \\
\hline
ECDF &  0.81 & $( 0.734,  0.886  )$ & $( 0.740,  0.890  )$ & $( 0.730,  0.880  )$ & $( 0.720,  0.870  )$ \\
\hline
Kernel 1D & 0.77  & $(  0.695,  0.838  )$ & $( 0.697,  0.840  )$ & $( 0.701,  0.843  )$ & $( 0.688,  0.833 )$  \\
\hline
Kernel 2D & 0.75  & $(  0.674,  0.807  )$ & $( 0.674,  0.807  )$ & $( 0.683,  0.816  )$ & $( 0.668,  0.804 )$  \\
\hline
MLE 1D &  0.78  & $( 0.707,  0.853  )$ & $( 0.709,  0.854  )$ & $(  0.705,  0.850 )$ & $( 0.701,  0.847  )$ \\
\hline
MLE 2D & 0.77   & $( 0.695,  0.845  )$ & $( 0.697,  0.846  )$ & $( 0.703,  0.853  )$ & $( 0.685,  0.840  )$ \\
\hline
SMLE 1D & 0.77  & $( 0.704,  0.844  )$ & $(  0.707,  0.847 )$ & $(   0.694,  0.835 )$ & $(  0.694,  0.835 )$ \\
\hline
SMLE 2D & 0.78  & $( 0.690,  0.830  )$ & $( 0.692,  0.832  )$ & $(  0.690,  0.830 )$ & $( 0.682,  0.824  )$ \\
\hline
\end{tabular}
\caption{\small Simulated data: Estimators and $95\%$ bootstrap confidence intervals.}
\label{table:EstCISAS}
\end{center}
\end{table}

\pagebreak

\subsection{Real data}\label{Real data}

In this section we study the data set presented in \cite{V96}, which contains 72 lesion scores obtained using both a clinical scheme without a dermoscope ($X$ Test), and a dermoscopic scoring scheme ($Y$ Test). Their main interest is to assess the information provided by the use of the dermoscope. Here, we analyse the subset of 51 non-diseased patients (diagnosed using a biopsy) and compare the nonparametric inferences for $\theta$ obtained using the estimators described in the introduction of this section.  It is important to note that the population correlation coefficient of this sample is $0.794$, which suggests that the entries are correlated.

\begin{figure}[!ht]
\begin{center}
\psfig{figure=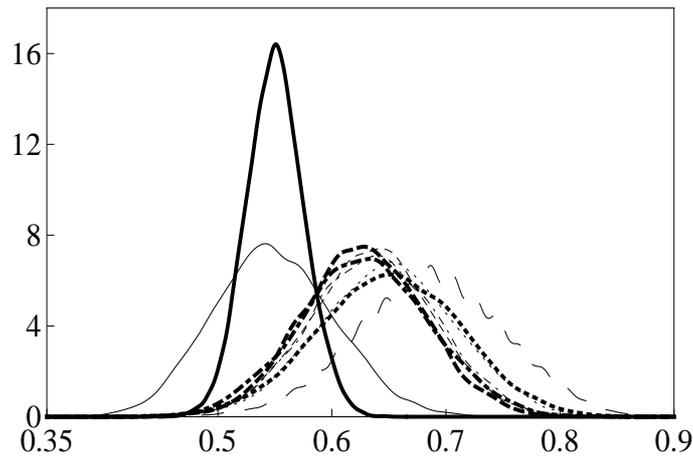, height=6.0cm, width=9cm}
\end{center}
\caption{\small Melanoma data: bootstrap distributions of the estimators. Independent (solid line), Paired (solid bold line), ECDF (long-dashed line), Kernel 1D (dashed line), Kernel 2D (dashed bold line), MLE 1D (dotted line), MLE 2D (dotted bold line), SMLE 1D (dotted-dashed line), SMLE 2D (dotted-dashed bold line). }
\label{fig:melanoma}
\end{figure}

Table \ref{table:EstCI} shows point estimators and four types of bootstrap confidence intervals of $\theta$. Figure \ref{fig:melanoma} shows the bootstrap distributions of the estimators of $\theta$. We can note a discrepancy of the point estimators under the assumptions of dependence and independence of the tests. Interval inference is also different; in the cases where pairing and dependence are not considered we can observe that the value $\theta=0.5$ is included in some of the bootstrap confidence intervals, leading to different conclusions about the relationship of the tests. This is in line with the conclusions in \cite{RS12} and emphasises the importance of the dependence and pairing assumptions.

\begin{table}[h!]
\begin{center}
\begin{tabular}[h]{ c c c c c c }
\hline
Estimator & $\hat{\theta}$ & Normal & Basic & Percentile & BCa \\
\hline
Independent & 0.55 & $( 0.469,  0.678 )$ & $( 0.467,  0.672 )$ & $( 0.450,  0.656 )$ & $( 0.474,  0.691 )$ \\
\hline
Paired & 0.55 & $( 0.498,  0.597 )$ & $( 0.497,  0.596 )$ & $( 0.501,  0.601 )$ & $( 0.499,  0.598 )$ \\
\hline
ECDF & 0.69 & $( 0.559,  0.813 )$ & $( 0.569,  0.823 )$ & $( 0.549,  0.804 )$ & $( 0.529,  0.784 )$ \\
\hline
Kernel 1D & 0.64 & $( 0.525,  0.737 )$ & $( 0.525,  0.738 )$ & $( 0.528,  0.741 )$ & $( 0.519,  0.732 )$ \\
\hline
Kernel 2D & 0.62 & $( 0.514,  0.720 )$ & $( 0.511,  0.719 )$ & $( 0.526,  0.733 )$ & $( 0.512,  0.718 )$ \\
\hline
MLE 1D & 0.65 & $( 0.543,  0.776 )$ & $( 0.544,  0.776 )$ & $( 0.532,  0.765 )$ & $( 0.537,  0.768 )$ \\
\hline
MLE 2D & 0.65 & $( 0.523,  0.769 )$ & $( 0.524,  0.772 )$ & $( 0.524,  0.772 )$ & $( 0.513,  0.762 )$ \\
\hline
SMLE 1D& 0.64 & $( 0.538,  0.756 )$ & $( 0.539,  0.757 )$ & $( 0.527,  0.744 )$ & $( 0.533,  0.749 ) $ \\
\hline
SMLE 2D& 0.63 & $( 0.519,  0.746 )$ & $( 0.523,  0.748 )$ & $( 0.511,  0.736 )$ & $( 0.512,  0.737 ) $ \\
\hline
\end{tabular}
\caption{\small Melanoma data: Estimators and $95\%$ bootstrap confidence intervals.}
\label{table:EstCI}
\end{center}
\end{table}

\pagebreak

\section{Discussion}

We introduced two classes of nonparametric estimators of $\theta = P(X<Y)$ for the case of paired, possibly dependent, observations. The proposed estimators avoid making assumptions on the distribution and the dependence structure of $(X,Y)$ which are implicitly considered by estimating nonparametrically either the joint distribution of $(X,Y)$ or the distribution of the difference $Z=Y-X$. We proved that the combination of the proposed approach with several nonparametric distribution estimators produce estimators of $\theta$ with appealing asymptotic properties. In addition, we have shown that confidence intervals for $\theta$, based on these estimators, can be obtained using bootstrap methods that are easy to implement using already existing R packages. The nonparametric distribution estimators explored in the context of Estimator I perform similarly. They are also comparable in terms of their ease of implementation and the required CPU usage. In the context of Estimator II, we empirically found that the estimators of $\theta$ based on smooth distribution estimators exhibit a better performance than those based on discrete distribution estimators such as the empirical distribution. The example presented in Section \ref{Real data} show that not accounting for dependence between $X$ and $Y$ may lead to opposite conclusions about $\theta=0.5$, and consequently about the relationship between these variables.


\end{document}